\documentclass[pdflatex,sn-mathphys-num,iicol]{sn-jnl}
\usepackage{multirow}
\usepackage{amsthm}
\usepackage{mathrsfs}
\usepackage[title]{appendix}
\usepackage{geometry}
\usepackage{textcomp}
\usepackage{tabularx}
\usepackage{manyfoot}
\usepackage{booktabs}
\usepackage{algorithm}
\usepackage{algorithmicx}
\usepackage{algpseudocode}
\usepackage{listings}
\usepackage{cite}
\usepackage{amsmath,amssymb,amsfonts}
\usepackage{graphicx, multicol}
\usepackage{xcolor}
\usepackage{physics}
\usepackage{lmodern}
\usepackage{kpfonts}
\usepackage{subcaption}
\usepackage{comment}
\usepackage{array}

\newcolumntype{M}[1]{>{\centering\arraybackslash}m{#1}}

\begin{document}

\title[Optical Quantum Mixed-State Reconstruction With Multiple Deep Learning Approaches]{Optical Quantum Mixed-State Reconstruction With Multiple Deep Learning Approaches}

\author*[1]{\fnm{Nhan} Trong \sur{Luu}}\email{luutn@ctu.edu.vn}
\author[2]{\fnm{Tuyen} Quang \sur{Nguyen}}\email{tuyen.q.nguyen@student.uts.edu.au}
\author[3]{\fnm{Duong} Trung \sur{Luu}}\email{luutd@ctu.edu.vn}
\author[4]{\sur{Truong} Cong \fnm{Thang}}\email{thang@u-aizu.ac.jp}

\affil*[1]{\orgdiv{College of Communication and Information Technology}, \orgname{Can Tho University}, \orgaddress{\state{Can Tho}, \country{Vietnam}}}

\affil[2]{\orgdiv{School of Computer Science}, \orgname{University of Technology Sydney}, \orgaddress{Ultimo, NSW 2007, \city{Sydney}, \country{Australia}}}

\affil[3]{\orgdiv{Center of Digital Transformation and Communication}, \orgname{Can Tho University}, \orgaddress{\city{Can Tho}, \country{Vietnam}}}

\affil[4]{\orgdiv{Department of Computer Science and Engineering}, \orgname{The University of Aizu}, \orgaddress{\city{Aizuwakamatsu}, \state{Fukushima}, \country{Japan}}}

\abstract{Quantum state tomography is a crucial technique for characterizing the state of a quantum system, which is essential for many applications in quantum technologies. In recent years, there has been growing interest in leveraging neural networks to enhance the efficiency and accuracy of quantum state tomography. However, versatile methods that are broadly applicable across diverse reconstruction scenarios remain relatively underexplored. In this paper, we present two neural network-based reconstruction approaches for both pure and mixed quantum state tomography: Restricted Feature Based Neural Network and Mixed States Neural Network. By leveraging class information during reconstruction, we are able to achieve state-of-the-art performance of tomography for both pure and mixed quantum states.}

\keywords{Quantum state tomography; quantum reconstruction; convolutional neural network; deep neural network}

\maketitle

\section{Introduction}
Recently many advancements of small quantum systems have been presented. These systems offer potential applications in various fields such as quantum information processing and computation~\citep{feynman2018simulating, montanaro2016quantum, wendin2017quantum}, quantum chemistry simulations~\citep{georgescu2014quantum, kandala2017hardware, childs2018toward}, secure communication~\citep{kimble2008quantum, yin2020entanglement}, among others notable works~\citep{georgescu2014quantum, you2011atomic, gu2017microwave}. In recent breakthroughs, the successful demonstration of a small quantum computer accomplishing a computational task in a fraction of the time expected on a classical supercomputer~\citep{arute2019quantum} highlights the remarkable speedup achievable through quantum computing. This acceleration is attributable in part to the exponentially vast state space available for storing and manipulating information in quantum systems~\citep{nielsen2010quantum, caves2004physical, havlivcek2019supervised}. Nonetheless, while the expansive state space offers opportunities, it also presents challenges in accurately characterizing and describing these systems.

Quantum state tomography (QST)~\citep{d2003quantum, liu2005tomographic, lvovsky2009continuous} aims to ascertain the unknown quantum state by conducting measurements on a finite set of identical copies of the system. If the state is characterized by the density matrix $\rho$, residing in a $d$-dimensional Hilbert space, approximately $O(d/\varepsilon)$ copies are necessary to achieve an estimate of $\rho$ with an error (measured as total variation distance) less than $\varepsilon$~\citep{o2016efficient}. This underscores the considerable resource demands of QST for large-scale systems.

In addition, machine learning techniques have been employed in QST, showing promising outcomes~\citep{carleo2017solving, xu2018neural, torlai2018neural}. Specifically, generative models~\citep{carrasquilla2019reconstructing, lloyd2018quantum, kieferova2017tomography}, often in the form of restricted Boltzmann machines (RBMs), have emerged as effective Ansätze with minimal parameters to depict a quantum state and understand the probability distribution of anticipated outputs~\citep{glasser2018neural, carleo2017solving, tiunov2020experimental}. Additionally, deep neural networks have been utilized in QST~\citep{ahmed2021classification, ahmed2021quantum}, enabling physicists to harness the swift advancements in machine learning methodologies. Still, a fully versatile deep learning (DL) method applicable across multiple scenarios remains relatively underexplored. By leveraging prior knowledge of the reconstructing quantum state, we are able to achieve QST by introduce two neural networks (NN) architectures: 
\begin{itemize}
    \item \textit{Restricted Feature-Based Neural Network (RFB-Net):} RFB-Net is an extended version of QST-NN~\citep{ahmed2021classification} to classify quantum states into specific classes and reconstruct their features under given constraints, thereby framing QST as a multitask learning problem.
    \item \textit{Mixed States Neural Network (MS-NN):} In contrast, MS-NN is derived from the generator of the QST conditional GAN (QST-CGAN)~\citep{ahmed2021quantum} and utilizes assisted state referencing to increase convergence speed, along with an improved \emph{DensityMatrix} process to extend its reconstruction range to include mixed states.
    \item Both proposed model are proven to be significantly outperforming traditional approaches~\citep{teo2011quantum, ahmed2021quantum} in reconstructing a diverse amount of quantum states' density matrices on both homodyne and heterodyne measurements.
\end{itemize}
Within the best of our knowledge, this is the first work that:
\begin{itemize}
    \item tackling quantum tomography using deep learning approaches that are versatile to both pure and mixed quantum states while being applicable to both homodyne and heterodyne measurement,
    \item providing a DL-based multitask leaning approach (RFB-Net) for quantum discrimination and reconstruction in parallel, allowing us to solve on both problem at the same time,
    \item generalizing the traditional Cholesky decomposition to both pure and mixed quantum states.
\end{itemize}

\section{Background}

In this section we will discuss some background knowledge related to our research, including knowledge related to QST and usage of neural network (NN) methods in QST.

\subsection{Quantum state tomography}

At its core, quantum state tomography (QST) addresses an inverse problem of quantum information science: given access to many identical copies of an unknown quantum system, how can one infer the underlying state from the collected measurement statistics? The standard workflow consists of repeatedly probing copies of the system with a chosen set of measurements and then applying statistical estimators to recover an object that fully captures the state, which is most often expressed through its density matrix $\rho$.

Reconstruction targets in QST generally fall into two regimes, namely pure states and mixed states. A pure state corresponds to a density matrix that is Hermitian, positive semi-definite, of unit trace, and admits a rank-one factorization through the state vector itself, namely
\begin{equation}
\rho = |\psi\rangle \langle \psi|
\end{equation}
For a mixed state, the density matrix instead takes the form of a statistical ensemble:
\begin{equation}
\rho = \sum_i p_i |\psi_i\rangle \langle \psi_i|
\end{equation}
with $p_i$ being the classical probability assigned to each constituent pure state $|\psi_i\rangle$. Because no single ket can fully encode a statistical mixture~\citep{zwiebach2022mastering}, working directly with density matrices yields a single, consistent framework that accommodates both regimes simultaneously.

For pure-state reconstruction, prior works~\citep{ahmed2021classification, ahmed2021quantum} have parameterized the density matrix through a Cholesky-style factorization,
\begin{equation}
\rho = LL^{\top}
\end{equation}
in which $L$ is constrained to be lower triangular with complex off-diagonal entries and real-valued diagonal entries. Such a parameterization is convenient because it intrinsically enforces both Hermiticity and positivity of the resulting matrix.

A widely adopted route to obtaining $\rho$ proceeds through single-shot readouts based on a positive operator-valued measure (POVM) $\{O_i\}_{i=1}^m$. Each measurement outcome contributes one entry to a data vector $\mathbf{d}$, where the $i^{\text{th}}$ entry corresponds to the expected value of the matching POVM element on the unknown state:
\[
d_i = \operatorname{Tr}(O_i\,\rho), \quad \text{for } i = 1, 2, \dots, m.
\]
Stacking all entries yields the column vector
\[
\mathbf{d} =
\begin{pmatrix}
d_1 \\
d_2 \\
\vdots \\
d_m
\end{pmatrix}
=
\begin{pmatrix}
\operatorname{Tr}(O_1\,\rho) \\
\operatorname{Tr}(O_2\,\rho) \\
\vdots \\
\operatorname{Tr}(O_m\,\rho)
\end{pmatrix}.
\]
This formulation casts reconstruction as a linear inverse problem~\citep{tarantola2005inverse, aster2018parameter, shen2016optimized}:
\begin{equation}
\mathbf{d} = A\,\rho_f
\label{inversion_eq}
\end{equation}
where the sensing matrix $A$ is induced by the chosen measurement operators and $\rho_f$ is a vectorized representation of the density matrix. Whether Equation~\ref{inversion_eq} can be inverted at all depends on the structure of the measurement set: configurations that fully constrain the state and thus admit unique inversion are referred to as informationally complete (IC) measurements~\citep{d2004informationally}. In the $N$-dimensional Hilbert-space setting, on the order of $N^2$ POVM elements are typically required to reach informational completeness, and each element may need to be sampled many times so that the empirical statistics converge to the true expectation values. Nonetheless, when additional structure is known in advance, such as the state being low-rank or having a sparsity pattern, the effective measurement budget can be considerably trimmed by exploiting that structure.

Once the measurement vector $\mathbf{d}$ has been collected, recovering $\rho$ becomes a statistical estimation task that admits a variety of solution strategies. Among the most widely used are linear inversion~\citep{qi2013quantum}, maximum-likelihood estimation (MLE)~\citep{banaszek1999maximum, lvovsky2004iterative}, and Bayesian inference~\citep{blume2010optimal, granade2016practical}. Linear inversion is conceptually simple, but it is highly sensitive to statistical fluctuations in $\mathbf{d}$ and to poorly conditioned sensing matrices $A$~\citep{shen2016optimized}, which in practice can yield unphysical reconstructions such as density matrices with negative diagonal entries. Likelihood-based and Bayesian estimators are therefore generally favored, since they explicitly incorporate the probabilistic structure of the measurements. In particular, MLE seeks the estimator $\rho'$ that maximizes the data likelihood:
\begin{equation}
L(\rho' \mid \textbf{d}) = \prod_i [\text{tr}(\rho' O_i)]^{d_i}
\end{equation}

When the outcomes $d_i$ are continuous rather than discrete, the likelihood is typically evaluated after binning the measurement domain~\citep{silva2018quadrature}. A related strategy bypasses the likelihood altogether and instead fits $\rho$ by minimizing the mean squared error between the model-predicted and measured statistics~\citep{smolin2012efficient}.

While MLE always returns a valid density matrix, it offers no native mechanism for assessing the uncertainty of its estimate. Subsequent studies have furthermore shown that MLE need not be optimal and is in fact inadmissible under certain figures of merit, including fidelity, mean-squared error, and relative entropy~\citep{ferrie2018maximum}. Bayesian estimation addresses the uncertainty issue by placing a prior $\pi(\rho)$ on the space of density matrices~\citep{blume2010optimal, granade2016practical}. The initial prior $\pi_0(\rho)$ is commonly selected to be either uniform or otherwise weakly informative, and the data are then incorporated via Bayes' rule with the likelihood $L(\rho' \mid d)$, producing the posterior $\pi_f(\rho) \propto L(\rho \mid d) \pi_0(\rho)$. The Bayesian point estimate is taken as the posterior mean:
\begin{equation}
\rho_\mu = \int \rho \pi_f(\rho) d\rho
\end{equation}

Beyond these standard estimators, several refinements have been proposed for likelihood maximization and density-matrix estimation, including diluted MLE~\citep{vrehavcek2007diluted}, compressed-sensing (CS) approaches~\citep{gross2010quantum, ahn2019adaptive}, and projected-gradient-descent (PGD) schemes~\citep{gonccalves2016projected, bolduc2017projected}. The motivation for CS comes from a parameter-counting perspective: only roughly $O(rN)$ samples are needed to identify a rank-$r$ density matrix~\citep{kalev2015quantum}. This makes CS particularly attractive for low-rank scenarios such as pure states perturbed by local noise, and recent adaptive variants~\citep{ahn2019adaptive, ahn2019adaptive2} go one step further by requiring only the Hilbert-space dimension, removing the need to know or estimate the rank $r$ ahead of time.

Another line of work avoids parameterizing $\rho$ in full generality and instead relies on compact ansätze that capture the state using far fewer than the generic $N^2 - 1$ free parameters. Examples include matrix-product-state (MPS) tomography~\citep{cramer2010efficient, lanyon2017efficient}, tensor-network (TN) tomography~\citep{chabuda2020tensor}, and permutationally invariant tomography~\citep{toth2010permutationally}, the last of which exploits exchange symmetries of the density matrix to make mixed-state reconstruction tractable. Other methods build in explicit noise assumptions, such as additive Gaussian readout noise~\citep{smolin2012efficient}, although such tailored techniques tend not to transfer cleanly to other noise regimes.

PGD~\citep{gonccalves2016projected, bolduc2017projected} provides yet another alternative, in which gradient-based optimization is applied to a cost function comparing predicted and observed measurement statistics, with iterates projected back onto the space of valid density matrices. PGD has the practical advantage of converging rapidly toward the MLE solution in a wide range of regimes, including settings where the inverse problem is poorly conditioned.

\subsection{Neural network based approach in quantum state tomography}

Neural networks have recently gained prominence as a robust method for extracting compact representations from high-dimensional datasets~\citep{hinton2007learning, graves2013speech, lecun2015deep}. In the realm of experimental quantum science, these methods have been successfully employed for tasks such as classifying experimental snapshots~\citep{rem2019identifying, bohrdt2019classifying} and qubit readout~\citep{seif2018machine}. This data-driven methodology can also be extended to tomographic tasks and are showing promising results in QST. These methods either use neural networks as ansatz to predict measurement probabilities~\citep{tiunov2020experimental, toth2010permutationally, torlai2019integrating} or directly estimate the density matrix \( \rho \)~\citep{lohani2020machine}. Recent theoretical advances have shown that a generative model, specifically the restricted Boltzmann machine (RBM), can accurately reconstruct quantum states and observables using synthetic datasets produced by numerical algorithms~\citep{torlai2020machine}.

Noticeably,~\citep{jia2019quantum, torlai2018latent} had explored the possibility of applying RBM for mixed state tomography. To reconstruct a mixed state density matrix $\rho_S$, the approach mirrors method used for pure states. We construct a neural network with parameters $\theta$, and for fixed input bases $|v\rangle$ and $|v'\rangle$, the density operator is represented by the matrix elements $\rho(\theta, v, v')$, which are determined by the neural network. Consequently, the mapping from a neural network with parameters $\theta$ to a density operator can be expressed as:

\begin{equation}
\rho(\theta) = \sum_{v,v'} |v\rangle \rho(\theta, v, v') \langle v'|
\end{equation}

To achieve this, we employ the purification method for density operators. In this framework, the environment is modeled by additional hidden neurons $e_1, \ldots, e_m$, in addition to the hidden neurons $h_1, \ldots, h_l$. The purification $|\Psi_{SE}\rangle$ of $\rho_S$ is represented by the parameters of the neural network, which we still denote by $\theta$, such that:

\begin{equation}
|\Psi_{SE}\rangle = \sum_{v} \sum_{e} \Psi_{SE}(\theta, v, e) |v\rangle |e\rangle
\end{equation}

By tracing out the environmental degrees of freedom, the density operator is determined by the network parameters as:

\begin{equation}
\rho_S = \sum_{v,v'} \left( \sum_{e} \Psi_{SE}(\theta, v, e) \Psi^{*}_{SE}(\theta, v', e) \right) |v\rangle \langle v'|
\end{equation}

Despite these advances, a versatile solution for large-scale NN-based QST solution that can accommodate various reconstruction scenarios has not yet been thoroughly explored. We aim to address this gap in the following sections.

\section{Data and noise model}

In this section, we will discuss the employed quantum states used to gather measurements for our reconstruction experiments and introduced noise during benchmarking process.

\subsection{Experimental data}\label{dataset}
By the assistance of QuTiP~\citep{johansson2012qutip}, our generated QST dataset include (but not limited to) 10000 measurements of 32-qubit Coherent states, Fock states, Thermal states, Cat states, Num states~\citep{albert2018performance, michael2016new}, Binomial states~\citep{albert2018performance, michael2016new} and GKP states~\citep{albert2018performance, campagne2020quantum, gottesman2001encoding} (number of samples uniformly distributed between states), along with the original density matrices, labels and generated values based on certain constraints. We measured our synthetic quantum states in 2 type of setting, creating 2 versions of our dataset:
\begin{itemize}
    \item \textit{Homodyne measurement version:} We employed quadrature field measurements \citep{grandi2017experimental} defined by the generalized field quadrature operator:
    \begin{equation}
        \hat{x}{\theta} = \frac{1}{\sqrt{2}}\left(\hat{a}e^{-i\theta} + \hat{a}^\dagger e^{i\theta}\right),
        \label{eq:quadrature_operator}
    \end{equation}
    where $\hat{a}$ and $\hat{a}^\dagger$ denote the annihilation and creation operators, respectively, and $\theta$ specifies the measurement phase.
    The corresponding quadrature probability distribution for a quantum state $\rho$ is given by
    \begin{equation}
        p(x\theta|\rho) = \bra{x_\theta}\rho\ket{x_\theta},
        \label{eq:quadrature_prob}
    \end{equation}
    where $\ket{x_\theta}$ are the eigenstates of the quadrature operator $\hat{x}_\theta$.
    \item \textit{Heterodyne measurement version:} We utilized the Husimi Q-function POVM \citep{nielsen2010quantum} to obtain the measurement of all quantum states $\rho$:
    \begin{equation}
        Q(\rho, \beta) = \frac{1}{\pi}\bra{\beta}\rho\ket{\beta},
    \label{eq:q_func}
    \end{equation}
    with coherent state $\ket{\beta}$ of amplitude $\beta$.
\end{itemize}

Typically, these states exist within a Hilbert space that has infinite dimensions. However, we can create a limited representation of these states with finite dimensions by setting a threshold on their energy levels, which is why in our experiments a Hilbert-space cutoff of $N_c=32$ is used for all cases. To prevent artefacts resulting from truncation, the maximum photon number of the states is limited to 16 after the displacements are applied. Different types of quantum optical states used in the research are defined based on the original paper of QST-NN~\citep{ahmed2021classification}, including three well-known basic classes and four states from bosonic codes designed for quantum error correction. Other constraints for dataset generation can be found in Table \ref{tab:constraint}.

\begin{table}[t!]
    \centering
    \resizebox{\linewidth}{!}{
    \begin{tabular}{ll}
     \hline
     \bf Name & \bf Constraint\\
     \hline
     Fock state & $1\leq n_{photon} \leq 16$\\
     \hline
     Coherent state &$10^{-6}\leq \abs{\alpha}^2 \leq 3$\\
     \hline
     Thermal state & $1 \leq n_{th} \leq 16$\\
     \hline
     Cat state & $S \in [0,2], \abs{\alpha} \in [1,3], r \in [0, 2S+1]$\\
     \hline
     Num state & $\overline{n} \in \{1.56, 2.67, 2.77, 4.15, 4.34\}$\\
     \hline
     Binomial state & $2 \leq N \leq N_C/(S+1)-1$\\
     \hline
     GKP state & $n_1, n_2 \in \{-20, 20\}, \delta \in [0.2, 0.5]$\\
     \hline
    \end{tabular}}
    \caption{Constraints table for state's properties in dataset}
    \label{tab:constraint}
\end{table}

\subsubsection{Fock state}
\begin{figure}[t!]
    \centering
    \begin{subfigure}{\linewidth}
      \centering
      \includegraphics[width=\linewidth]{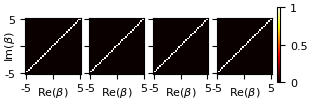}
      \caption{Homodyne}
    \end{subfigure}
    ~
    \begin{subfigure}{\linewidth}
      \centering
      \includegraphics[width=\linewidth]{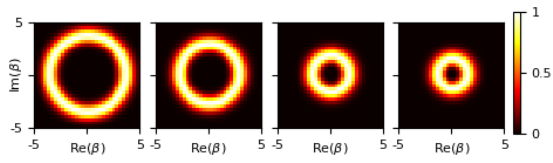}
      \caption{Heterodyne}
    \end{subfigure}
    \caption{Measurement samples of Fock state.}
    \label{fig:fock}
\end{figure}

Fock states form the number-state basis of a single bosonic mode and are simultaneous eigenstates of the photon-number operator~\citep{nielsen2010quantum, lvovsky2009continuous}, given by
\begin{equation}
\ket{\psi_{fock}} =\ket{n_{photon}}.
\end{equation}
A visualization of this state is shown in Figure \ref{fig:fock}.

\subsubsection{Coherent state}
\begin{figure}[t!]
    \centering
    \begin{subfigure}{\linewidth}
      \centering
      \includegraphics[width=\linewidth]{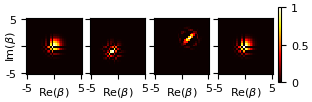}
      \caption{Homodyne}
    \end{subfigure}
    ~
    \begin{subfigure}{\linewidth}
      \centering
      \includegraphics[width=\linewidth]{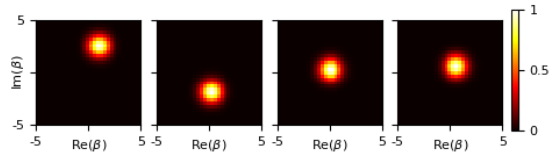}
      \caption{Heterodyne}
    \end{subfigure}
    \caption{Measurement samples of Coherent state}
    \label{fig:coherent}
\end{figure}

Coherent states are produced by applying a displacement operator to the vacuum, with the complex amplitude $\alpha$ specifying the displacement in phase space~\citep{nielsen2010quantum, lvovsky2009continuous},
\begin{equation}
\begin{aligned}
    \ket{\psi_{coherent}(\alpha)} &= \ket{\alpha}\\ 
    &= D(\alpha)\ket{0}\\
    D(\alpha) &= exp(\alpha\times a^{\dagger} - \alpha\times a)
\end{aligned}
\end{equation}
Here $D(\alpha)$ denotes the displacement operator, while $a$ and $a^{\dagger}$ act as the annihilation and creation operators of the bosonic mode, respectively. A visualization of this state is shown in Figure \ref{fig:coherent}.

\subsubsection{Thermal state}
\begin{figure}[t!]
    \centering
    \begin{subfigure}{\linewidth}
      \centering
      \includegraphics[width=\linewidth]{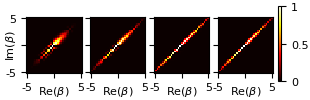}
      \caption{Homodyne }
    \end{subfigure}
    ~
    \begin{subfigure}{\linewidth}
      \centering
      \includegraphics[width=\linewidth]{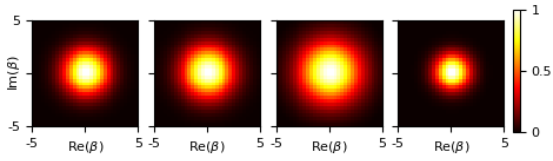}
      \caption{Heterodyne}
    \end{subfigure}
    \caption{Measurement samples of Thermal state}
    \label{fig:thermal}
\end{figure}

Thermal states describe a bosonic mode in thermal equilibrium with its environment, expressed as a diagonal mixture in the Fock basis whose populations follow a Bose-Einstein distribution~\citep{nielsen2010quantum, lvovsky2009continuous}:
\begin{equation}
\rho_{thermal}(n^{th}) = \sum_{n=0}^{N_c-1}\frac{1}{n^{th}+1}(\frac{n^{th}}{n^{th}+1})^n\ket{n}\bra{n}.
\end{equation}
A visualization of this state is shown in Figure \ref{fig:thermal}.

\subsubsection{Cat state}
\begin{figure}[t!]
    \centering
    \begin{subfigure}{\linewidth}
      \centering
      \includegraphics[width=\linewidth]{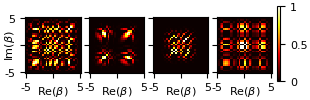}
      \caption{Homodyne }
    \end{subfigure}
    ~
    \begin{subfigure}{\linewidth}
      \centering
      \includegraphics[width=\linewidth]{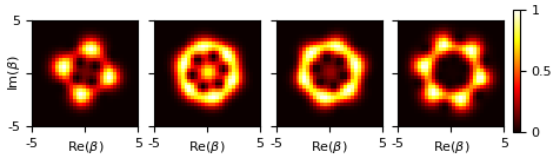}
      \caption{Heterodyne}
    \end{subfigure}
    \caption{Measurement samples of Cat state}
    \label{fig:cat}
\end{figure}

Cat states are bosonic-code states obtained from coherent-state superpositions, originally studied as even and odd coherent states~\citep{dodonov1974even} and later generalized in the bosonic-code framework~\citep{albert2018performance, michael2016new}. Up to a normalization factor $N$, they are constructed with projection operators $\Pi_r$ as
\begin{equation}
\begin{aligned}
    \ket{\psi_{cat}^{\mu}} &= \frac{1}{N}\Pi_{(S+1)\mu}\{\ket{\alpha e^{i\frac{\pi}{S+1}k}}\}^{2S+1}_{k=0},\\
    \Pi_r &= \sum^{\infty}_{m=0}\ket{2m(S+1)+r}\bra{2m(S+1)+r}.
\end{aligned}
\end{equation}
A visualization of this state is shown in Figure \ref{fig:cat}.

\subsubsection{Num state}

\begin{figure}[t!]
    \centering
    \begin{subfigure}{\linewidth}
      \centering
      \includegraphics[width=\linewidth]{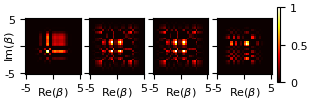}
      \caption{Homodyne }
    \end{subfigure}
    ~
    \begin{subfigure}{\linewidth}
      \centering
      \includegraphics[width=\linewidth]{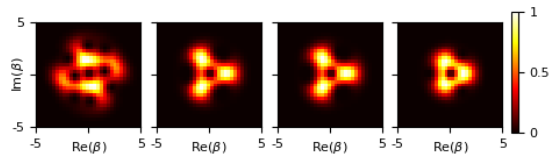}
      \caption{Heterodyne}
    \end{subfigure}
    \caption{Measurement samples of Num state}
\label{fig:num}
\end{figure}

Num states form a class of bosonic codes built from low-weight Fock-state superpositions, with their coefficients numerically tuned for quantum-error-correction performance~\citep{albert2018performance, michael2016new}, and indexed by the mean photon number $\overline{n}$. A visualization of this state is shown in Figure \ref{fig:num}.

\subsubsection{Binomial state}
\begin{figure}[t!]
    \centering
    \begin{subfigure}{\linewidth}
      \centering
      \includegraphics[width=\linewidth]{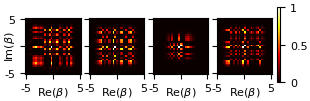}
      \caption{Homodyne }
    \end{subfigure}
    ~
    \begin{subfigure}{\linewidth}
      \centering
      \includegraphics[width=\linewidth]{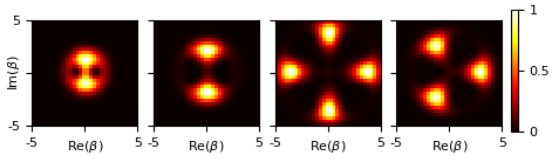}
      \caption{Heterodyne}
    \end{subfigure}
    \caption{Measurement samples of Binomial state}
\label{fig:bin}
\end{figure}

Binomial states are bosonic-code states whose Fock-basis amplitudes are weighted by binomial coefficients~\citep{albert2018performance, michael2016new}, taking the form
\begin{equation}
\ket{\psi_{binomial}^{\mu}} = \frac{1}{\sqrt{2^{N+1}}}\sum^{N+1}_{m=0}(-1)^{\mu m} \sqrt{\begin{pmatrix}
N+1\\
m
\end{pmatrix}} \ket{(S+1)m}.
\end{equation}
A visualization of this state is shown in Figure \ref{fig:bin}.

\subsubsection{GKP state}
\begin{figure}[t!]
    \centering
    \begin{subfigure}{\linewidth}
      \centering
      \includegraphics[width=\linewidth]{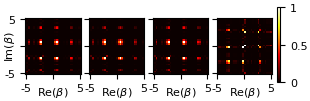}
      \caption{Homodyne }
    \end{subfigure}
    ~
    \begin{subfigure}{\linewidth}
      \centering
      \includegraphics[width=\linewidth]{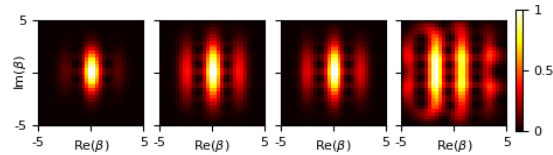}
      \caption{Heterodyne}
    \end{subfigure}
    \caption{Measurement samples of GKP state}
    \label{fig:gkp}
\end{figure}

The Gottesman-Kitaev-Preskill (GKP) encoding~\citep{gottesman2001encoding} embeds a logical qubit in the continuous-variable Hilbert space of a bosonic mode through a phase-space lattice structure. Since the idealized GKP states are non-normalizable, we adopt the finite-energy version~\citep{albert2018performance, campagne2020quantum} that truncates the lattice and applies a Gaussian envelope:
\begin{equation}
\begin{aligned}
    \ket{\psi_{GKP}^{\mu}} &= \sum_{\alpha \in K(\mu)} e^{-\delta^2|\alpha|^2}e^{-iRe[\alpha]Im[\alpha]}\ket{\alpha}, \\
    K(\mu) &= \sqrt{\frac{\pi}{2}}(2n_1+\mu) + i\sqrt{\frac{\pi}{2}}n_2,
\end{aligned}
\end{equation}
A visualization of this state is shown in Figure \ref{fig:gkp}.

\subsection{Experimented noise models}

Imperfections in a real tomography pipeline can enter at three logically distinct points. The first is at the preparation stage, where the target state itself may be imperfectly produced and what is actually generated is a perturbed density operator $\rho \rightarrow \rho_{\text{noisy}}$. The second is at the measurement stage, where calibration drift or imperfect implementation of the POVM elements causes the nominal operators to differ from the ones effectively realized in hardware, $\{ \mathcal{O}_i \} \rightarrow \{ \mathcal{O}_{\text{noisy},i} \}$. The third sits at the readout layer, where the recorded counts are themselves perturbed -- by amplifier noise, photon shot fluctuations, or downstream electronics -- so that the data vector observed by the reconstruction algorithm is $\mathbf{d} \rightarrow \mathbf{d}_{\text{noisy}}$ rather than the ideal one.

Errors of the first two kinds, jointly referred to as state-preparation and measurement (SPAM) errors, tend to be systematic and persistent, which makes them hard to remove by purely classical post-processing. A practical workaround that has gained traction in recent years is to delegate the correction itself to a learned model: a supervised neural network is trained to map noisy data back toward the clean ones, schematically $\mathbf{d}_{\text{noisy}} \rightarrow \text{DNN} \rightarrow \mathbf{d}$~\citep{palmieri2020experimental}. From this angle, the network plays the role of a learned denoising filter that does not need an explicit physical model of the underlying SPAM process.

A natural way to phrase the broader question -- whether neural-network reconstructors can stay reliable when confronted with noise models they have never seen -- is through the lens of \emph{domain adaptation}: the predictor must remain accurate even when its input distribution shifts away from the one it was trained on, while still mapping into the same target space.

Recent theoretical and empirical findings indicate that high accuracy on a single training distribution (in-distribution generalization) tends to carry over, at least partially, to related but unseen distributions (out-of-distribution generalization), without any retraining required \citep{miller2021accuracy}. More formally, the gap between expected losses under two input distributions admits an upper bound expressed through a statistical divergence between the two. Writing $\mathcal{D}_s$ and $\mathcal{D}_t$ for the source and target input distributions, and given a hypothesis $h$ together with a loss function $\ell$, one has \citep{ben2010theory}
\begin{equation}
\left| 
\mathbb{E}_{x \sim \mathcal{D}_s}[\ell(h(x))] 
- 
\mathbb{E}_{x \sim \mathcal{D}_t}[\ell(h(x))]
\right|
\;\le\;
\mathrm{TV}(\mathcal{D}_s, \mathcal{D}_t),
\label{eq:domain_bound}
\end{equation}
where $\mathrm{TV}(\mathcal{D}_s, \mathcal{D}_t)$ stands for the total-variation distance between the two distributions.

The practical implication is that a reconstructor exposed during training to a sufficiently broad family of synthetic noise channels can be expected to behave reasonably on real SPAM-corrupted data, as long as the gap between the synthetic and the experimental input distributions is not too large in total-variation terms. This is precisely the rationale behind augmenting our training data with explicit noise channels rather than training on idealized measurements alone. Within this paper, we use the noise families described in the rest of this section as a stress test, and report the resulting fidelities in Section \ref{noise_benchmark}.

\subsubsection{Mixed state noise}

Thermal occupation of unused modes, residual coupling to a bath, and other environmental factors cause the prepared state to drift away from a pure target. We capture this behavior in a coarse-grained way by mixing the ideal state with a randomly sampled density operator:
\begin{equation}
\rho_{\text{mixed}} = (1 - \zeta) \rho + \zeta \rho_{\text{random}},
\end{equation}
with mixing weight $\zeta \in [0, 0.5]$. Because $\rho_{\text{mixed}}$ -- not $\rho$ -- is what an experiment would actually emit, it is also $\rho_{\text{mixed}}$ that becomes the reconstruction target in this scenario.

\subsubsection{Photon loss noise}

In a lossy cavity, photons can leak out of the resonator during the time window between state preparation and measurement, leaving the post-decay state different from the one originally prepared. We capture this dissipative dynamic by propagating the initial density operator for a duration $\tau$ under the Lindblad-form master equation
\begin{equation}
\dot{\rho} = -\frac{i}{\hbar} [H, \rho] + \gamma \mathcal{L}[a] \rho
\end{equation}
in which $H = \hbar \omega a^{\dagger} a$ is the bare resonator Hamiltonian at frequency $\omega$, $\gamma$ sets the photon-loss rate, and the dissipator acts as $\mathcal{L}[a] \rho = a \rho a^{\dagger} - \frac{1}{2} a^{\dagger} a \rho - \frac{1}{2} \rho a^{\dagger} a$.

\subsubsection{Pepper noise}

Pepper noise is a one-sided variant of the classical salt-and-pepper corruption model and is meant to emulate isolated dropouts in the measurement signal, such as those caused by dead pixels in a sensor array or by missing samples in a data stream. Concretely, we apply it by randomly selecting a fraction of entries in each measurement and forcing those entries to zero.

\section{Architecture and training settings}

In this section, we will introduce the architectural design of NN-based models and some notable training settings used in our experiments.

\subsection{Restricted Feature Based Neural Network}
\subsubsection{Model implementation}

\begin{figure*}[t!]
\centerline{\includegraphics[width=\linewidth]{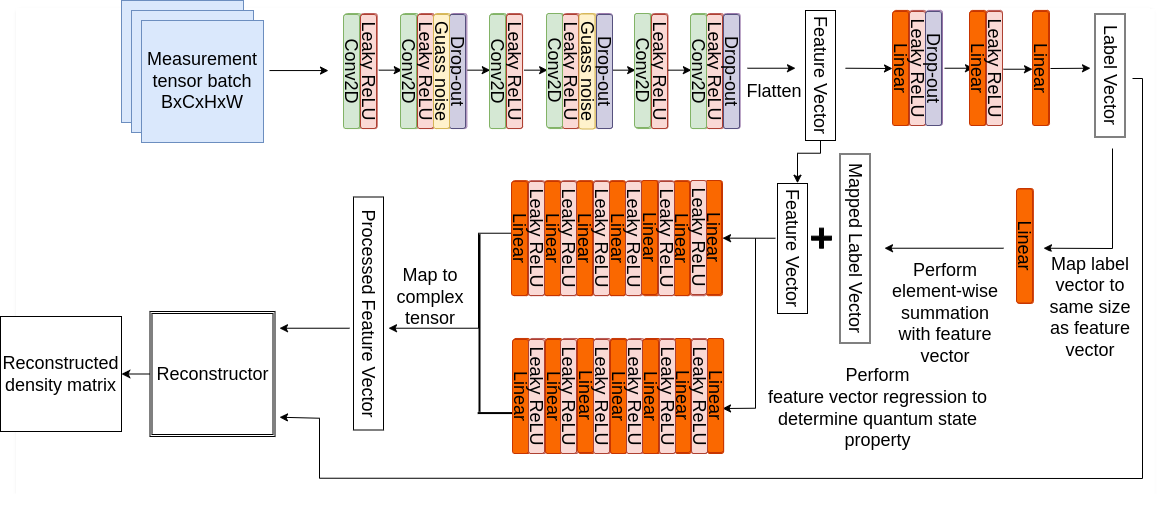}}
\caption{Architecture of RFB-Net.}
\label{fig:rfb-net}
\end{figure*}

\begin{table*}[t]
\centering
\caption{Layer specification of the proposed RFB-Net model.}
\resizebox{0.8\textwidth}{!}{\begin{tabular}{l l l l}
\hline
\textbf{Module} & \textbf{Layer} & \textbf{Output Dimension} & Activation\\
\hline
\multirow{13}{*}{Feature extractor} 
& Input & $B \times 1 \times H \times W$  & -- \\

& Conv2D ($3\times3$, no bias) & $B \times 32 \times (H{-}2) \times (W{-}2)$ & LeakyReLU ($\alpha=0.01$)\\

& Conv2D ($3\times3$, no bias) & $B \times 32 \times (H{-}4) \times (W{-}4)$ & LeakyReLU ($\alpha=0.01$)\\
& Gaussian noise ($\sigma=0.005$) & -- & --\\
& Dropout ($p=0.4$) & -- & --\\

& Conv2D ($3\times3$, stride $2$) & $B \times 64 \times \frac{H{-}6}{2} \times \frac{W{-}6}{2}$ & LeakyReLU ($\alpha=0.01$)\\

& Conv2D ($3\times3$) & $B \times 128 \times \frac{H{-}8}{2} \times \frac{W{-}8}{2}$ & LeakyReLU ($\alpha=0.01$)\\
& Gaussian noise ($\sigma=0.005$) & -- & -- \\
& Dropout ($p=0.4$) & -- & -- \\

& Conv2D ($3\times3$) & $B \times128 \times \frac{H{-}10}{2} \times \frac{W{-}10}{2}$ & LeakyReLU ($\alpha=0.01$)\\

& Conv2D ($3\times3$, stride $2$) & $B \times 64 \times \frac{H{-}12}{4} \times \frac{W{-}12}{4}$ & LeakyReLU ($\alpha=0.01$)\\
& Dropout ($p=0.4$) & -- & -- \\
\hline

Flattening & Flatten & $1024$ & --\\
\hline

\multirow{4}{*}{Real-valued head}
& Linear & $1024 \rightarrow 512$ & LeakyReLU ($\alpha=0.01$)\\
& Dropout ($p=0.4$) & -- & --\\
& Linear & $512 \rightarrow 256$ & --\\
& Linear & $256 \rightarrow 7$ & --\\
\hline

\multirow{10}{*}{Complex regressor}
& Real mapping & $7 \rightarrow 1024$ & -- \\
& Imag mapping & $7 \rightarrow 1024$ & --\\
& Feature fusion & $+$ flattened features & --\\
& Linear blocks (real) & $\begin{aligned}
    & 1024 \rightarrow 512 \rightarrow 512\\
    & \rightarrow 256 \rightarrow 256 \rightarrow 128 \rightarrow 3\\
\end{aligned}$ & LeakyReLU ($\alpha=0.1$)\\
& Linear blocks (imag) & $\begin{aligned}
    & 1024 \rightarrow 512 \rightarrow 512\\
    & \rightarrow 256 \rightarrow 256 \rightarrow 128 \rightarrow 3\\
\end{aligned}$ & LeakyReLU ($\alpha=0.1$) \\
& Output (real) & $\mathbb{R}^{3}$ & --\\
& Output (imag) & $\mathbb{R}^{3}$ & --\\
\hline

Model output & Final output & $(x,\; \mathbf{c} \in \mathbb{C}^{3})$ & --\\
\hline
\end{tabular}}
\label{tab:rfb_spec}
\end{table*}

We utilizes the PyTorch~\citep{paszke2019pytorch} framework to implement a deep convolutional neural network (CNN) similar to QST-NN. In our proposed architecture, we aim to extract relevant information from input measurements that are provided in the form of 32x32 density matrices. To accomplish this, we employ a series of 6 convolutional heads, which are interspersed with Gaussian noise, dropout, and leaky ReLU activation functions to extract meaningful features from the input.

Subsequently, we feed the extracted features into both a classification tail and a regression tail. The classification tail is responsible for predicting the label, while the regression tail computes three features of the state. To generate the final state prediction, we sum the predicted label output with the extracted features and utilize the regression tail to predict the three essential features of the state. 

These three features are then fed into a \emph{Reconstructor} module, which is an algorithm designed explicitly to reconstruct the state by taking 3-length value vectors as input and generating the predicted states. RFB-Net diagram can be found at Figure \ref{fig:rfb-net}, layer specification can be found at Table \ref{tab:rfb_spec} and details on how the \emph{Reconstructor} is implemented can be found at Algorithm \ref{alg:recon}.

\begin{algorithm}[t!]
\caption{Reconstructor algorithm implementation}
\label{alg:recon} 
\begin{algorithmic}
\Require Hilbert dimension $d$, labels $\mathbf{l}$, parameters $\mathbf{v}$
\Ensure Density matrix (or batch) $\rho$

\Function{Reconstruct}{$l, v$}
    \If{$l = 0$}
        \State \Return $\mathrm{Fock}(d, \mathrm{round}(\Re v_0))$
    \ElsIf{$l = 1$}
        \State \Return $\mathrm{Coherent}(d, v_0)$
    \ElsIf{$l = 2$}
        \State \Return $\mathrm{Thermal}(d, \Re v_0)$
    \ElsIf{$l = 3$}
        \State $i \gets \mathrm{clip}(\mathrm{round}(\Re v_0), -5, 4)$
        \State $\mu \gets \mathrm{clip}(\mathrm{round}(\Re v_1), -3, 2)$
        \State \Return $\mathrm{Num}(d, i, \mu)$
    \ElsIf{$l = 4$}
        \State $\begin{aligned}
            \text{\Return } \mathrm{Binomial}(& d, \mathrm{round}(\Re v_0),\\
        &\mathrm{round}(\Re v_1),\\
        &\mathrm{round}(\Re v_2))\\
        \end{aligned}$
    \ElsIf{$l = 5$}
        \State $\begin{aligned}
            \text{\Return } \mathrm{Cat}(& d, v_0,\\
        &\mathrm{round}(\Re v_1),\\
        &\mathrm{round}(\Re v_2))
        \end{aligned}$
    \Else
        \State \Return $\mathrm{GKP}(d, \Re v_0,
        \mathrm{round}(\Re v_1))$
    \EndIf
\EndFunction

\If{$\mathbf{l}$ is scalar}
    \State \Return $\mathrm{Reconstruct}(\mathbf{l}, \mathbf{v})$
\EndIf
\State $\rho \gets \{\mathrm{Reconstruct}(l_i, v_i)\}_{i=1}^{|\mathbf{l}|}$
\State \Return $\rho$
\end{algorithmic}
\end{algorithm}

\subsubsection{Model-specific training settings}

As noted from previous sections, the aim of this model is to solve quantum tomography as a multitask learning problem~\citep{zhang2021survey}, in which several complementary properties of the quantum state are inferred simultaneously in order to facilitate accurate reconstruction. Formally, given some input $\mathbf{x}$, the model learns a shared representation:
\begin{equation}
    \mathbf{h} = f_{\theta}(\mathbf{x}),
\end{equation}
from which multiple task-specific predictors $g_{\theta_{k}}$ for $K$ amount of task are derived:
\begin{equation}
    \mathbf{z}_{k} = g_{\theta_{k}}(\mathbf{h}), 
    \quad k = 1,\dots,K.
\end{equation}
Training is then performed by minimizing a joint objective function,
\begin{equation}
\min_{\theta}\mathcal{L}(\hat{\mathbf{z}}, \mathbf{z}) =
\sum_{k=1}^{K} \lambda_{k} \, \mathcal{L}_{k}(\hat{\mathbf{z}}_{k}, \mathbf{z}_{k}),
\end{equation}
where $\mathcal{L}_{k}$ are task-specific loss functions and $\lambda_{k}$ probabilistically balance the contribution of each task. Based on these theoretical foundation, we train our model using a multi-objective loss function with cross-entropy for the classification target and the sum of mean absolute error (MAE) for the real and imaginary parts of the regression output \emph{F} as a feature vector:
\begin{equation}
\begin{aligned}
    \mathcal{L}(y, \hat{y}, F, \hat{F}) = & \frac{1}{N} \sum_{n=1}^{N} l_n(y, \hat{y}, F, \hat{F}), \\
    l_n(y, \hat{y}, F, \hat{F}) = & -\sum_{c=1}^{C} w_c \log\left( \frac{\exp{y_{n,c}}}{\sum_{i=1}^{C} \exp{y_{n,i}}} \right) \hat{y}_{n,c}\\ 
    & +\abs{Re(F_{n,c}) - Re(\hat{F}_{n,c})} \\
    & + \abs{Im(F_{n,c}) - Im(\hat{F}_{n,c})}
\end{aligned} 
\end{equation}
where \( y \) and \( \hat{y} \) denote the input and target respectively, \( w \) the class weight, \( C \) the number of classes, and \( N \) the batch size.

\subsection{Mixed State Neural Network}

\subsubsection{Model implementation}

\begin{figure*}[t]
\centerline{\includegraphics[width=\linewidth]{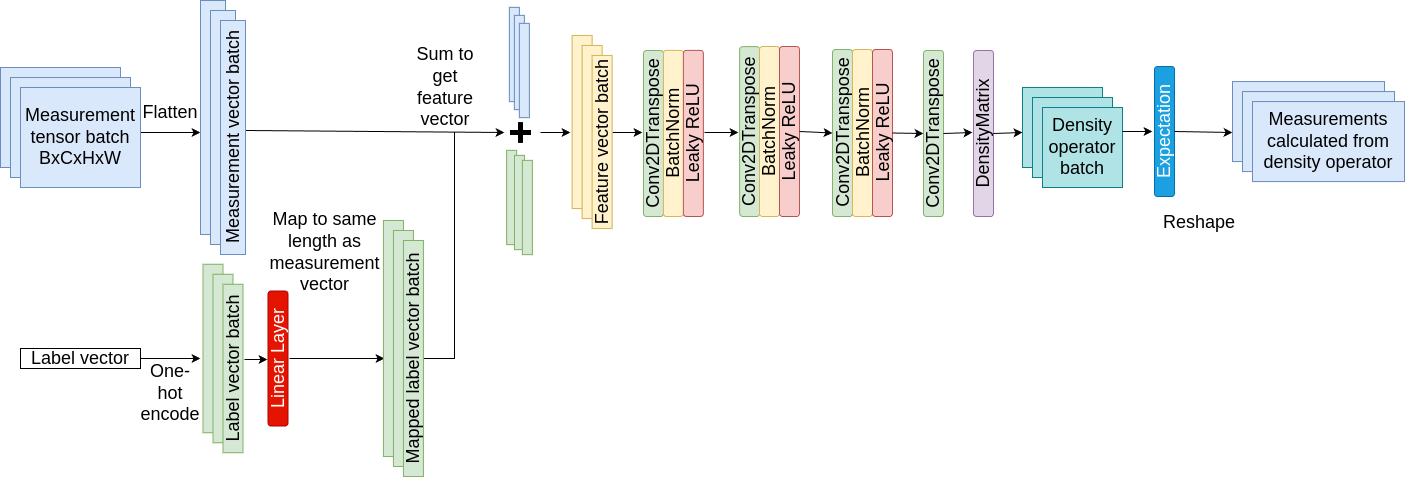}}
\caption{Architecture of MS-NN.}
\label{fig:msnn}
\end{figure*}

\begin{table*}[t]
\centering
\caption{Layer specification of the proposed MS-NN model.}
\resizebox{0.8\textwidth}{!}{\begin{tabular}{llll}
\hline
\textbf{Block} & \textbf{Layer / Operation} & \textbf{Output Dimension} & \textbf{Activation} \\
\hline

\multirow{3}{*}{Inputs}
& Measurement input & $H \times H$ & -- \\
& Operator input & $H \times H \times (2H^2)$ & -- \\
& Label input & $1$ & -- \\
\hline

Measurement encoding
& Reshape & $1 \times H^2$ & -- \\
\hline

Label encoding
& One-hot encoding & $N_{\mathrm{class}}$ & -- \\
& Dense ($N_{\mathrm{class}} \rightarrow H^2$, no bias) & $H^2$ & -- \\
& Addition with measurement & $H^2$ & -- \\
\hline

\multirow{8}{*}{Multi-scale decoder}
& ExpandDims & $1 \times H^2 \times 1$ & -- \\
& Conv2DTranspose ($4\times4$, 256 ch., stride 1) & $4 \times (H^2{-}3) \times 256$ & -- \\
& BatchNorm & Same as above & LeakyReLU ($\alpha=0.01$) \\
\cmidrule(lr){2-4}

& Conv2DTranspose ($4\times4$, 128 ch., stride 1) & $7 \times (H^2{-}6) \times 128$ & -- \\
& BatchNorm & Same as above & LeakyReLU ($\alpha=0.01$)\\
\cmidrule(lr){2-4}

& Conv2DTranspose ($4\times4$, 64 ch., stride 2) & $14 \times \cdot \times 64$ & -- \\
& BatchNorm & Same as above & LeakyReLU ($\alpha=0.01$)\\
\cmidrule(lr){2-4}

& Conv2DTranspose ($2\times2$, 2 ch., stride 2) & $H \times H \times 2$ & -- \\
\hline

\multirow{3}{*}{Physical projection}
& Cholesky decomposition & $H \times H \times 2$ & -- \\
& DensityMatrix& $H \times H$ (complex) & -- \\
& Trace normalization & $H \times H$ (complex) & -- \\
\hline

\multirow{3}{*}{Measurement head}
& Operator conversion & $H \times H$ (complex) & -- \\
& Expectation values & $H^2$ & -- \\
& Reshape & $H \times H$ & -- \\
\hline

Model outputs
& Density matrix $\rho$ & $H \times H$ (complex) & -- \\
& Predicted expectations & $H \times H$ & -- \\
\hline
\end{tabular}}
\label{tab:msnn_spec}
\end{table*}

Building upon the seminal work of QST-CGAN~\citep{ahmed2021quantum}, we derived MS-NN from QST-CGAN's generator~\citep{ahmed2021quantum}, utilizing TensorFlow~\citep{abadi2016tensorflow} framework. Our proposed model is designed to accommodate a batch of measurements in the form of 32x32 density matrices, as well as label vectors representing the target state. To facilitate efficient processing, we have implemented a series of pre-processing steps, which include flattening the measurements to a 1024-length vector and one-hot encoding the label to a 7-length vector.

Subsequently, we feed the label vector through a dense layer, which enables us to encode it into a 1024-length vector that can be mapped to the same size as the measurements. By adding the resulting vector and the flattened measurements, we generate a final feature that captures the essence of the underlying quantum state. This final feature is then fed through a series of convolutional layers, batch normalization, and leaky ReLU activation, resulting in a batch of 2x32x32 tensor, where the two channels correspond to the real and imaginary components of the quantum state. Diagram of MS-NN implementation can be found at Figure \ref{fig:msnn} and layer specification can be found in Table \ref{tab:msnn_spec}.

\subsubsection{Cholesky decomposition generalization for both pure and mixed quantum states}

Following the generation of the complex tensor batch, we proceed to further process the data by passing it through a \emph{DensityMatrix} layer, which is responsible for normalizing the quantum state in accordance with specific constraints. Subsequently, we pass the normalized state through an \emph{Expectation} layer, which is designed to measure the state by calculating the expected values of various observables.

\begin{figure}[t!]
    \centerline{\includegraphics[width=\linewidth]{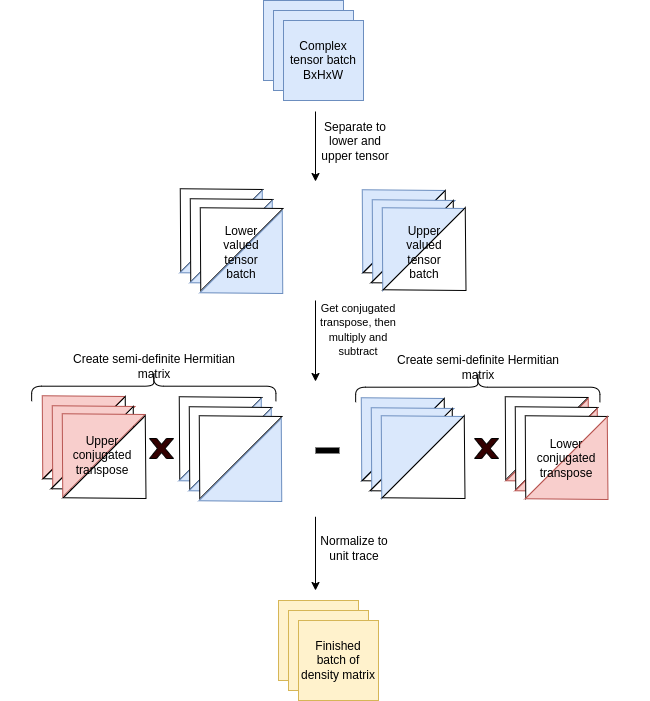}}
    \caption{Graphical demonstration of the improved \emph{DensityMatrix} process.}
    \label{fig:dm}
\end{figure}

Unlike the original Cholesky decomposition utilized in prior works \citep{teo2011quantum, ahmed2021quantum} which only utilizes the lower half $L \in \mathbb{C}^{n \times n}$ of the result to obtain a Hermitian matrix as density matrix:
\[
    \rho = \frac{L L^{\dagger}}{\operatorname{Tr}\!\left(L L^{\dagger}\right)},
\]
our novel approach also simulate a Cholesky decomposition~\citep{banaszek1999maximum} to the upper half $U \in \mathbb{C}^{n \times n}$ of the result, thereby optimizing utilized model parameters $\theta$ in the process: 
\begin{equation}
\begin{aligned}
    & \begin{aligned}
        \rho & = \left(LL^{\top} - UU^{\top}\right) \oslash diag(LL^{\top} - UU^{\top})\\
        & \text{ where $L, U \in \theta$} \\
    \end{aligned}\\
    &\forall \mathbf{z}\in \mathbb{C}^{n}
    \begin{cases}
      Pr(\mathbf{z^*}\rho\mathbf{z}\geq 0) \in (0, 1)\\
      Pr(\mathbf{z^*}\rho\mathbf{z}\leq 0) \in (0, 1)
    \end{cases} 
\end{aligned}
\end{equation}
where $\oslash$ denote the Hadamard division and $diag()$ is the diagonal matrix extract operation. By subtracting the two obtained matrices and normalizing the result, we are able to generate a density operator that encompasses positive semi-definite $Pr(\mathbf{z^*}\rho\mathbf{z}\geq 0)$, negative semi-definite $Pr(\mathbf{z^*}\rho\mathbf{z}\leq 0) \in (0, 1)$, and even indefinite outcomes. This approach is particularly relevant in our situation where the dataset contains odd cat states~\citep{dodonov1974even} created through the subtraction of two coherent states $\ket{\alpha}$ and $\ket{-\alpha}$:
\begin{equation}
\ket{\psi_{cat}^{\mu}} = \frac{1}{\sqrt{2\left(1-e^{-2\abs{\alpha}^2}\right)}}\left(\ket{\alpha} - \ket{-\alpha}\right).
\end{equation}
A more graphical demonstration of the improved \emph{DensityMatrix} version can be found in Figure \ref{fig:dm}.

\subsubsection{Model-specific training settings}

While the proposed MS-NN is largely inspired by the conditional-GAN design adopted in prior quantum tomography studies \citep{ahmed2021quantum, ahmed2021classification}, its underlying mechanism for quantum state reconstruction is more closely aligned with an auto-encoder–like architecture. In contrast to adversarial learning, where a generator is optimized against a discriminator, MS-NN directly learns a structured latent-to-physical mapping that reconstructs a valid quantum state from measurement data and conditioning variables.

In a standard auto-encoder \citep{li2023comprehensive}, an input $\mathbf{x} \in \mathcal{X}$ is first mapped to a low-dimensional latent representation $\mathbf{z} \in \mathcal{Z}$ via an encoder $f_{\theta}(\cdot)$, and subsequently reconstructed through a decoder $g_{\phi}(\cdot)$:
\begin{equation}
\mathbf{z} = f_{\theta}(\mathbf{x}), 
\qquad 
\hat{\mathbf{x}} = g_{\phi}(\mathbf{z}),
\label{eq:autoencoder}
\end{equation}
where the model parameters $(\theta, \phi)$ are optimized to minimize a reconstruction loss:
\begin{equation}
\mathcal{L}_{\mathrm{AE}} = \lVert \mathbf{x} - \hat{\mathbf{x}} \rVert.
\label{eq:ae_loss}
\end{equation}
Analogously, MS-NN encodes the measurement outcomes and conditional labels into an intermediate latent representation, which is subsequently decoded into a structured matrix representation that parameterizes a physical quantum state. From this perspective, MS-NN can be interpreted as a physics-informed auto-encoder, where the decoder is constrained by quantum mechanical principles and the reconstruction problem can be solved using traditional regression-like objective function. 

Based on these characteristic, model's loss function $\mathcal{L}$ is designed to be a linear combination of MAE between the true state measurement $\mathcal{M}( \hat{\rho})$ (where $\mathcal{M}$ is a generic measurement procedure, specified type of POVM used in each dataset can be found in Section \ref{dataset}) and the predicted state measurement $\mathcal{M}( \rho)$ with an $\alpha$ weighted sum ($\alpha=100$) of MAE for the real component and imaginary component of the density operator as:
\begin{equation}
    \begin{aligned}
        \mathcal{L}(\rho, \hat{\rho}, \beta) & = \frac{1}{N}\sum_{i=1}^{N}\abs{\mathcal{M}(\rho_i) - \mathcal{M}( \hat{\rho_i})} \\
        & + \alpha\abs{Re(\rho_i) - Re(\hat{\rho_i})}\\
        &+ \alpha\abs{Im(\rho_i) - Im(\hat{\rho_i})}\\
    \end{aligned}
\end{equation}
where the loss $\mathcal{L}$ is averaged over $N$ minibatch and $\ket{\beta}$ is a coherent state centered at point $\beta$ in a given phase space and $\beta \in \mathbb{C}$. Real and imaginary part of the quantum state is probabilistically emphasized in training since they hold greater value in reconstruction process than the measurement itself.
 
\subsection{Accuracy metrics and other settings}
In all of our experiment, we will be using Uhlmann fidelity \citep{nielsen2010quantum, ahmed2021quantum} to measure the similarity between two quantum states. Uhlmann fidelity is a key concept in quantum information theory and plays a critical role in many quantum information protocols, including quantum error correction, quantum teleportation, and quantum cryptography. Given two density matrices $\rho$ and $\sigma$, the fidelity is defined by : 
\begin{equation}
F(\rho, \sigma) = \left(Tr\sqrt{\sqrt{\rho}\sigma\sqrt{\rho}}\right)^2
\end{equation}
where if $F(\rho, \sigma) == 1$, the two states are identical. If the fidelity is $0 \leq F(\rho, \sigma) < 1$ , the two states differ, and the degree of difference is quantified by the fidelity value. 

All tested model are trained on the synthetic dataset by employing SGDR with an initial learning rate of $lr = 0.1 \times (\text{batch size}/256)$ using training practices from \citet{luu2025hybrid}, combined with Cosine Annealing scheduling method and warm restart from $lr/10$ \citep{loshchilov2016sgdr}, along with Nesterov momentum set to $nesterov=0.9$. The experimentation of both models are conducted with a batch size of 256 states per forward inference on a single NVIDIA GeForce RTX 3090 GPU. 

\section{Results}

\subsection{Performance comparison with prior works}

\begin{figure*}[t]
    \centerline{\includegraphics[width=\linewidth]{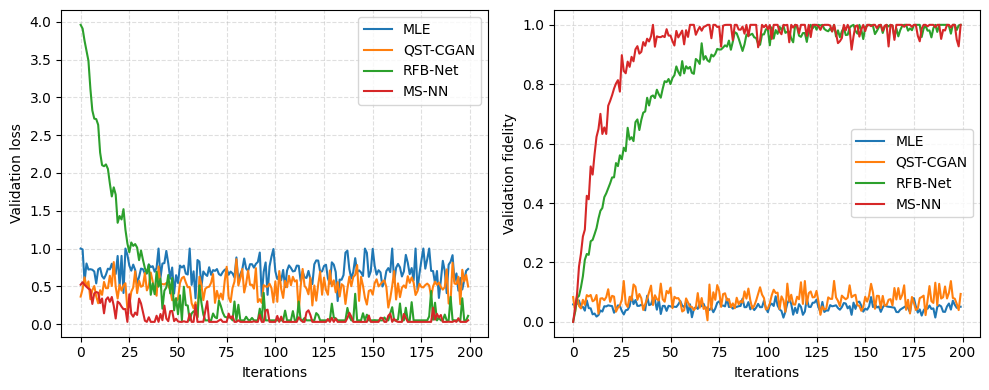}}
    \caption{Performance comparison figure between well known benchmarked NN-based method with our proposed models within 1 training-validation cycle, averaged over both homodyne and heterodyne dataset.}
    \label{fig:perf_compare}
\end{figure*}

\begin{table*}[t!]
    \centering
    \resizebox{\textwidth}{!}{
    \begin{tabular}{cccccc}
    \hline
    \multirow{2}{*}{\shortstack{Benchmarked\\model}} & \multirow{2}{*}{\shortstack{Parameter\\count}} & \multirow{2}{*}{\shortstack{Forward-pass\\memory cost (MB)}} & \multirow{2}{*}{\shortstack{Processing\\speed (ms)}} & \multicolumn{2}{c}{\shortstack{Fidelity per measurement\\type (higher is better)}} \\
    \cmidrule(lr){5-6}
     & & & & Homodyne & Heterodyne\\
    \hline
    MLE~\citep{teo2011quantum} & 1024 & 376.18~$\pm$~0.01 & 2.143~$\pm$~1.171 & 0.0843~$\pm$~0.0327 & 0.0914~$\pm$~0.0121\\
    \hline
    QST-CGAN~\citep{ahmed2021quantum} & 523K & 1970.65~$\pm$~0.03 & 30.02~$\pm$~5.268 & 0.1576~$\pm$~0.0241 & 0.1714~$\pm$~0.0121\\
    \hline
    \textbf{RFB-Net (Ours)} & 3.03M & 3102.18~$\pm$~0.02 & 75.21~$\pm$~9.101 & \textbf{0.9221~$\pm$~0.0421} & \textbf{0.9532~$\pm$~0.0573}\\
    \hline
    \textbf{MS-NN (Ours)} & 4.86M & 3890.11~$\pm$~0.03 & 87.94~$\pm$~10.03 & \underline{0.8901~$\pm$~0.0350} & \underline{0.9041~$\pm$~0.0424}\\
    \hline
    \end{tabular}}
    \caption{Average validation fidelity with standard deviation of investigated models on generated dataset averaged over 4 training-validation cycle. Best performing model is highlighted in bold and second best is underlined. We also compared other factor such as parameter count, memory usage in simulated device during forward pass (in megabyte) and processing speed on simulated device (in millisecond) per measurement batch of 32.}
    \label{tab:perf_compare}
\end{table*}

Table~\ref{tab:perf_compare} and Figure~\ref{fig:perf_compare} summarizes the performance of different reconstruction models: traditional maximum likelihood estimation (MLE) \citep{teo2011quantum}, QST-CGAN \citep{ahmed2021classification} and our proposed 2 model, evaluated on a generated dataset averaged over 4 training-validation cycles. The comparison includes both reconstruction accuracy, measured by fidelity under different measurement types, and computational efficiency metrics, including parameter count, forward-pass memory usage, and processing speed.

While MLE had served as a meaningful approach in various researches \citep{ahmed2021classification, ahmed2021quantum}, it only achieving moderate fidelity under both homodyne and heterodyne measurements, demonstrating an inferior performance when training scene contain a significantly more diverse reconstruction data.

Among neural-network-based methods, QST-CGAN~\citep{ahmed2021classification} slightly improve fidelity measurement over MLE due to a stronger representation capability of DNN models. Nevertheless, its reconstruction accuracy remains significantly lower than that of the proposed models, indicating limited generalization capability using traditional Cholesky decomposition method.

The proposed RFB-Net achieves the highest fidelity across both homodyne and heterodyne measurements, outperforming all baselines by a clear margin. Notably, both proposed methods improves fidelity by more than 4.5 times relative to QST-CGAN and by around 10 times compared to MLE, when reconstructing both homodyne and heterodyne data, demonstrating the effectiveness of multitask approach in capturing measurement-dependent quantum state features.

In terms of computational efficiency, RFB-Net strikes a favorable balance between model complexity and runtime performance. Although it has a higher reconstruction fidelity than MS-NN, it requires significantly less memory and computation than the MS-NN model, which exhibits comparable fidelity but at substantially higher computational cost. Specifically, MS-NN incurs nearly 800MB in forward-pass memory usage and more than $2\times$ slower processing speed, making it less suitable for practical deployment. Still, as seen from Figure~\ref{fig:perf_compare}, the proposed MS-NN demonstrates rapid and stable convergence thanks to large representation capability without having to multitask like RFB-Net. The validation loss decreases sharply within the first few tens of iterations and remains consistently low thereafter, with significantly reduced fluctuations. The fast fidelity convergence suggests that our architecture effectively captures quantum state features across different resolutions of the Hilbert space.

\subsection{Benchmarking with noise}\label{noise_benchmark}

\begin{figure*}[t]
    \begin{subfigure}{.5\textwidth}
      \begin{multicols}{3}
          \centering
          \includegraphics[width=.28\textwidth]{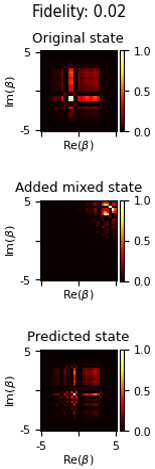}\\
          \includegraphics[width=.28\textwidth]{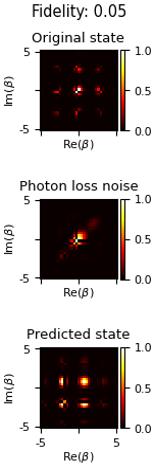}\\
          \includegraphics[width=.28\textwidth]{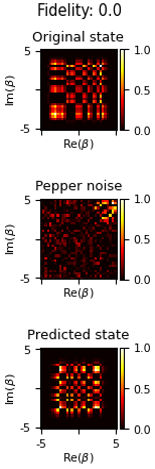}
      \end{multicols}
      \caption{RFB-Net homodyne reconstruction samples.}
    \end{subfigure}%
    ~
    \begin{subfigure}{.5\textwidth}
      \begin{multicols}{3}
          \centering
          \includegraphics[width=.28\textwidth]{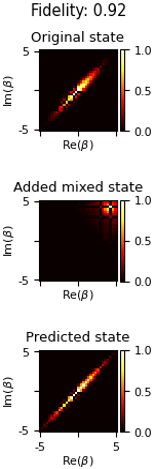}\\
          \includegraphics[width=.28\textwidth]{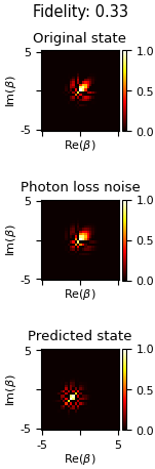}\\
          \includegraphics[width=.28\textwidth]{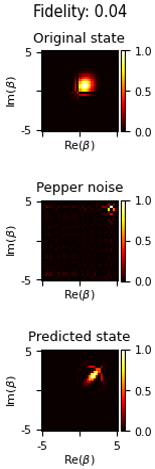}
      \end{multicols}
      \caption{MS-NN homodyne reconstruction samples.}
    \end{subfigure}
    ~
    \begin{subfigure}{.5\textwidth}
      \begin{multicols}{3}
          \centering
          \includegraphics[width=.28\textwidth]{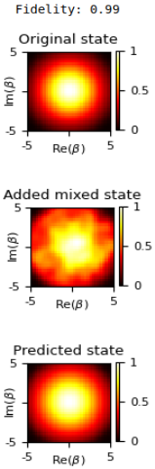}\\
          \includegraphics[width=.28\textwidth]{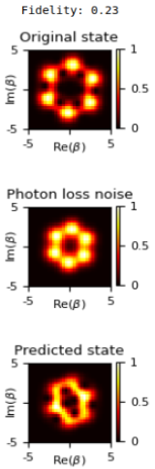}\\
          \includegraphics[width=.28\textwidth]{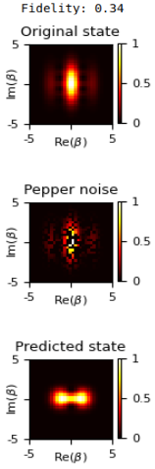}
      \end{multicols}
      \caption{RFB-Net heterodyne reconstruction samples.}
    \end{subfigure}%
    ~
    \begin{subfigure}{.5\textwidth}
      \begin{multicols}{3}
          \centering
          \includegraphics[width=.28\textwidth]{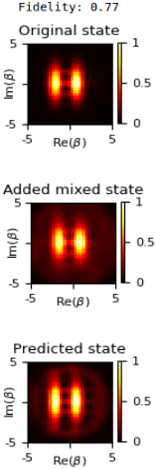}\\
          \includegraphics[width=.28\textwidth]{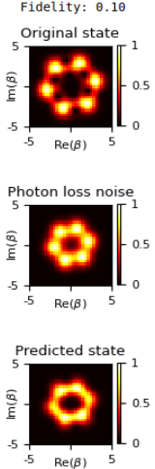}\\
          \includegraphics[width=.28\textwidth]{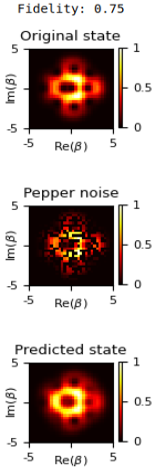}
      \end{multicols}
      \caption{MS-NN heterodyne reconstruction samples.}
    \end{subfigure}
    \caption{Samples of noise-added input measurements and untrained reconstruction results of RFB-Net and MS-NN.}
    \label{fig:noise_bench}
\end{figure*}

\begin{table*}[t!]
    \centering
    \resizebox{\textwidth}{!}{
    \begin{tabular}{cc*{6}{c}}
    \hline
    \multirow{3}{*}{\bf \shortstack{Benchmarking\\scenario}} & \multirow{3}{*}{\bf\shortstack{Model\\name}} & \multicolumn{2}{c}{\bf Mixed state noise} & \multicolumn{2}{c}{\bf Photon loss noise} & \multicolumn{2}{c}{\bf Pepper noise} \\
    \cmidrule(lr){3-4}\cmidrule(lr){5-6}\cmidrule(lr){7-8}
    & & \bf Homodyne & \bf Heterodyne & \bf Homodyne & \bf Heterodyne & \bf Homodyne & \bf Heterodyne\\
    \hline
    \multirow{3}{*}{\shortstack{Without added\\noise training}} & RFB-Net & 0.27~$\pm$~0.05 & 0.80~$\pm$~0.02 & 0.35~$\pm$~0.13 & 0.51~$\pm$~0.08& 0.14~$\pm$~0.07 & 0.29~$\pm$~0.11\\
    \cmidrule(lr){2-8}
    & MS-NN & 0.21~$\pm$~0.05 & 0.48~$\pm$~0.08 & 0.38~$\pm$~0.02 & 0.47~$\pm$~0.03 & 0.22~$\pm$~0.11 & 0.61~$\pm$~0.07\\
    \hline
    \multirow{3}{*}{\shortstack{With added\\noise training}} & RFB-Net & \underline{0.85~$\pm$~0.11} & \underline{0.87~$\pm$~0.09} & \underline{0.87~$\pm$~0.08} & \underline{0.83~$\pm$~0.11}& \underline{0.90~$\pm$~0.10} & \underline{0.89~$\pm$~0.09}\\
    \cmidrule(lr){2-8}
    & MS-NN & \textbf{0.92~$\pm$~0.05} & \textbf{0.95~$\pm$~0.09} & \textbf{0.93~$\pm$~0.05} & \textbf{0.90~$\pm$~0.02}& \textbf{0.91~$\pm$~0.06} & \textbf{0.94~$\pm$~0.07}\\
    \hline
    \end{tabular}}
    \caption{Average validation fidelity with standard deviation of our models on different noise type, averaged over 4 training and validation cycle. Best performing model is highlighted in bold and second best is underlined.}
    \label{tab:noise_bench}
\end{table*}

\subsubsection{Performance without Noise-Aware Training}

Upper half of Table~\ref{tab:noise_bench} summarizes the average validation fidelity (with standard deviation over four training-validation cycles) for different benchmarking scenarios when the models are trained without explicit noise augmentation. Across all noise types, both RFB-Net and the proposed MS-NN exhibit limited robustness, with fidelities strongly dependent on the measurement scheme and noise characteristics. Some visualization of the reconstruction result for both homodyne and heterodyne measurement data can be found in Figure~\ref{fig:noise_bench}.

Under mixed-state noise, MS-NN slightly outperforms RFB-Net in both homodyne and heterodyne settings, achieving average fidelities of $0.21$ and $0.48$, respectively. Although this improvement indicates better inductive bias in MS-NN, the absolute performance remains relatively low, suggesting that neither model generalizes well to unseen noise distributions when trained only on idealized data.

A similar trend is observed for photon loss noise. While RFB-Net shows moderate performance in the heterodyne case, both models struggle under homodyne measurements, reflecting the sensitivity of photon loss to measurement incompleteness. For pepper noise, MS-NN again demonstrates improved stability compared to RFB-Net, particularly in the heterodyne configuration. Nevertheless, the overall degradation in fidelity across all noise types highlights the inherent limitations of noise-agnostic training in practical quantum tomography scenarios, especially when dealing with large changes in input distribution of noisy homodyne measurement.

\subsubsection{Performance with Noise-Aware Training}

As seen from the lower half of Table~\ref{tab:noise_bench}, when noise-aware training is incorporated, a substantial performance improvement is observed for both architectures. In this setting, the models are trained using measurement data augmented with representative noise processes, enabling them to learn noise-resilient reconstruction strategies.

RFB-Net benefits significantly from noise-aware training, achieving fidelities above $0.85$ across most scenarios. However, the proposed MS-NN consistently achieves the highest performance, reaching near-unity fidelities in all tested noise types and measurement configurations. In particular, MS-NN attains fidelities of $0.92$ and $0.95$ for mixed-state noise under homodyne and heterodyne measurements, respectively, and maintains similarly high accuracy for photon loss and pepper noise. These results demonstrate that MS-NN adapts more effectively to noisy measurement statistics and exhibits superior generalization thanks to its large size and generalization capability.

\section{Ablation studies}

\subsection{Model-specific components}

\begin{table}[t!]
    \centering
    \resizebox{\linewidth}{!}{
    \begin{tabular}{cccc}
    \hline
    \multirow{2}{*}{\bf Model} & \multirow{2}{*}{\bf Removed component(s)} & \multicolumn{2}{c}{\bf \shortstack{Fidelity per measurement\\type (higher is better)}} \\
    \cmidrule(lr){3-4}
     & & \bf Homodyne & \bf Heterodyne\\
    \hline
    \multirow{4}{*}{RFB-Net} & All Gaussian noise layers & 0.91~$\pm$~0.02 & 0.93~$\pm$~0.06\\
    \cmidrule(lr){2-4}
    & All Dropout layers & 0.91~$\pm$~0.03 & 0.93~$\pm$~0.03\\
    \cmidrule(lr){2-4}
    & All Dropout and Gaussian & 0.89~$\pm$~0.07 & 0.90~$\pm$~0.04\\
    \hline
    MS-NN & All Batch Norm layers & 0.82~$\pm$~0.06 & 0.85~$\pm$~0.08\\
    \hline
    \end{tabular}}
    \caption{Ablation studies when removing certain operations from the model, averaged over 4 training and validation cycle.}
    \label{tab:abl_comp}
\end{table}

Table~\ref{tab:abl_comp} reports an ablation study evaluating the impact of noise injection and dropout regularization on reconstruction fidelity for both RFB-Net and the proposed MS-NN, under homodyne and heterodyne measurement schemes. The results are averaged over multiple runs, with standard deviations reflecting training stability.

For RFB-Net, removing all Gaussian noise layers leads to a noticeable degradation in performance, particularly under homodyne measurements, where the average fidelity drops to $0.91$. Although heterodyne fidelity remains relatively high, the increased variance indicates reduced robustness. Similarly, removing all dropout layers causes a comparable performance decline, suggesting that regularization plays a crucial role in preventing overfitting to specific measurement statistics. When both Gaussian noise and dropout are removed, the degradation becomes more pronounced, confirming that these components contribute complementary regularization effects. In contrast, MS-NN exhibits a different sensitivity profile. Removing batch normalization layers results in a significant decrease in fidelity for both homodyne and heterodyne measurements, with average fidelities of $0.82$ and $0.85$, respectively.

\subsection{Loss statistic balancing}

\begin{table}[t!]
    \centering
    \resizebox{\linewidth}{!}{
    \begin{tabular}{ccccc}
    \hline
    \multirow{2}{*}{\bf \shortstack{Model}} & \multicolumn{2}{c}{\bf Loss component weighting} & \multicolumn{2}{c}{\bf \shortstack{Fidelity per measurement\\type (higher is better)}} \\
    \cmidrule(lr){2-3}\cmidrule(lr){4-5}
     & \bf Classification & \bf Reconstruction & \bf Homodyne & \bf Heterodyne\\
    \hline
    \multirow{4}{*}{RFB-Net} & 1.0 & 1.0 & 0.92~$\pm$~0.04 & 0.95~$\pm$~0.05\\
    \cmidrule(lr){2-5}
     & 0.5 & 0.5 & 0.90~$\pm$~0.03 \textcolor{red}{($\downarrow$2.17\%)} & 0.91~$\pm$~0.01 \textcolor{red}{($\downarrow$4.21\%)}\\
    \cmidrule(lr){2-5}
     & 1.0 & 2.0 & 0.70~$\pm$~0.06 \textcolor{red}{($\downarrow$23.9\%)} & 0.72~$\pm$~0.05 \textcolor{red}{($\downarrow$24.2\%)}\\
    \hline
    & \bf Measurement & \bf Density matrix & \bf Homodyne & \bf Heterodyne\\
    \cmidrule(lr){2-5}
    \multirow{1}{*}{MS-NN} & 1.0 & 100.0 & 0.89~$\pm$~0.04 & 0.90~$\pm$~0.04\\
    \cmidrule(lr){2-5}
     & 1.0 & 1.0 & 0.22~$\pm$~0.17 \textcolor{red}{($\downarrow$75.3\%)} & 0.26~$\pm$~0.15 \textcolor{red}{($\downarrow$71.1\%)}\\
    \hline
    \end{tabular}}
    \caption{Ablation studies with different loss component statistic weighting, averaged over 4 training and validation cycle. The first result row of each model is the default setting used in our main results section. Fidelity reduction percentage in comparison with main results is denoted in red.}
    \label{tab:abl_loss}
\end{table}

Table~\ref{tab:abl_loss} presents an ablation study examining the effect of different loss component weightings for both RFB-Net and the proposed MS-NN. All fidelities are averaged over 4 independent training and validation cycles, with the first row for each model corresponding to the default configuration used in the main results.

For RFB-Net, the default configuration assigns equal weights to the classification and reconstruction losses and achieves fidelities of $0.92 \pm 0.04$ for homodyne measurements and $0.95 \pm 0.05$ for heterodyne measurements. When both loss components are uniformly reduced to $0.5$, the performance drops slightly to $0.90 \pm 0.03$ (homodyne) and $0.91 \pm 0.01$ (heterodyne), corresponding to relative decreases of $2.17\%$ and $4.21\%$, respectively. Although modest, this degradation suggests that uniformly scaling down the loss magnitude weakens the effective gradient signal, which slows optimization and leads to slightly worse convergence.

A much stronger effect is observed when the reconstruction objective is weighted more heavily than the classification objective. Setting the loss ratio to $(1.0, 2.0)$ results in fidelities of only $0.70 \pm 0.06$ for homodyne and $0.72 \pm 0.05$ for heterodyne measurements, corresponding to substantial decreases of $23.9\%$ and $24.2\%$, respectively. In absolute terms, this represents fidelity drops of $0.22$ and $0.23$ compared to the baseline configuration. These results highlight the sequential dependency between the two objectives: accurate state classification provides a meaningful prior for reconstruction. When the model prioritizes reconstruction before reliably identifying the quantum state class, it tends to converge to physically inconsistent solutions, resulting in significantly lower fidelity.

A similar sensitivity to loss balancing is observed for MS-NN, although the underlying mechanism differs. Under the default configuration, where the density-matrix loss is scaled by a factor of $100$, the model achieves fidelities of $0.89 \pm 0.04$ (homodyne) and $0.90 \pm 0.04$ (heterodyne). When both loss components are set to equal weights $(1.0, 1.0)$, the performance collapses dramatically to $0.22 \pm 0.17$ and $0.26 \pm 0.15$, corresponding to fidelity reductions of $75.3\%$ and $71.1\%$, respectively. In absolute terms, the average fidelity decreases by approximately $0.67$ for homodyne and $0.64$ for heterodyne measurements.

Furthermore, the standard deviation increases significantly under the equal-weight setting (from approximately $0.04$ to $0.15$-$0.17$), indicating unstable training dynamics and inconsistent convergence across runs. This behavior reflects the importance of the density-matrix-based objective in MS-NN. Because expectation-value matching acts as the physically grounded optimization target, reducing its relative influence weakens the constraint that enforces quantum-state consistency. Consequently, the model becomes more susceptible to ambiguous measurement statistics, where different quantum states can produce similar geometric measurement distributions, leading to unreliable reconstructions.

\section{Discussions and limitations}

Although both proposed architectures achieve substantial improvements over existing approaches, they exhibit different trade-offs that may influence model selection in practical deployments. RFB-Net consistently achieves the highest reconstruction fidelity, reaching $0.9221 \pm 0.0421$ and $0.9532 \pm 0.0573$ under homodyne and heterodyne measurements respectively, while maintaining lower computational cost than MS-NN. In contrast, MS-NN demonstrates faster convergence and improved robustness under noise-aware training scenarios, as shown in Table~\ref{tab:noise_bench}, where it achieves fidelities above $0.90$ across all evaluated noise types. However, this performance advantage comes at the expense of higher computational requirements, including approximately $25\%$ higher memory consumption and slower inference speed. Consequently, RFB-Net may be preferable in resource-constrained environments or when maximal reconstruction accuracy is required, whereas MS-NN may be advantageous in scenarios where robustness to noisy measurements and rapid convergence are prioritized.

It should be noted that the experiments in this study rely primarily on synthetically generated quantum measurement datasets. While theoretical results from domain adaptation suggest that models trained on sufficiently diverse synthetic distributions can generalize to real experimental data, further validation using physical quantum optical experiments will be necessary to confirm the practical applicability of the proposed methods.

\section{Conclusion}

This research paper explored the application of neural networks in quantum tomography. We propose two neural‑network architectures for quantum state reconstruction: the Restricted Feature‑Based Network (RFB‑Net), which performs direct state estimation using only measurement data, and the Mixed‑States Neural Network (MS‑NN), which leverages prior knowledge of the target state to achieve accelerated convergence. The results demonstrate that neural network architectures, when trained on appropriate datasets, can produce accurate and efficient reconstructions of quantum states. Furthermore, the approach presented in this paper is scalable and can be extended to large-scale quantum systems. It is clear that the combination of neural networks and quantum tomography has great potential for advancing the field of quantum information processing, and we anticipate that this research will inspire further exploration and development of these methods in the future.

\subsection*{Acknowledgment}
The authors received no funding for this research.

\subsection*{Author Contributions}
Nhan Trong Luu and Truong Cong Thang conceptualized the study, performed the data analysis and wrote the original draft. Nhan Trong Luu and Nguyen Quang Tuyen implemented the experiments. Duong Trung Luu, Nguyen Quang Tuyen assisted with the experiments and manuscript revision. All authors reviewed and approved the final manuscript.

\subsection*{Data availability}
Source code is publicly available at \url{https://github.com/luutn2002/uni-qst}.

\subsection*{Declarations}

\subsubsection*{Conflict of interest} 
The authors have no competing interests to declare that are relevant to the content of this article.

\subsubsection*{Ethical Approval} 
All applicable international, national, and/or institutional guidelines for the care and use of animals were followed.

\nocite{*}
\bibliography{sn-bibliography}

\end{document}